\documentclass[12pt]{article}

\newcommand{\bea}{\begin{eqnarray}}
\newcommand{\eea}{\end{eqnarray}}

\begin{document}

\title{\bf A simple collinear limit of scattering amplitudes at strong coupling}

\author{Gang Yang}
\date{}

\maketitle

\centerline{\it Centre for Research in String Theory}
\centerline{\it Department of Physics, Queen Mary, University of
London}
\centerline{\it Mile End Road, London, E1 4NS, United Kingdom}

\begin{abstract}

Collinear limit usually provides strong constraints for scattering
amplitudes. At strong coupling, collinear limit of the amplitudes in
${\cal N}$=4 SYM is related to the large mass limit of the
corresponding $Y$ system.
In this paper, we consider a special case in which all mass
parameters are taken to be large, which corresponds to a
multi-double-collinear limit in which a $n$-side polygon becomes
pentagons.
This limit provides a useful constraint for the amplitudes, in
particular can be used to fix the periods part for the case of $4K$
gluons, which is the last missing piece of full amplitudes.

\end{abstract}

\section{Introduction}

Using AdS/CFT duality, the problem of calculating scattering
amplitudes at strong coupling in ${\cal N}$=4 super Yang-Mills
theory is related to a geometric problem of computing the area of
minimal surface in $AdS_5$ background ending on a polygonal
light-like contour  \cite{AM}. At weak coupling, the duality between
amplitudes and null polygonal Wilson loops was also found
\cite{{Drummond123}, {Brandhuber:2007yx}, {Drummond3}}. This duality
revealed a remarkable hidden symmetry of planar scattering
amplitudes, the so-called dual superconformal symmetry \cite{AM,
Drummond3, {Drummond:2008vq}, {Berkovits:2008ic}, {Beisert:2008iq}}.
This is not a conventional conformal symmetry of Lagrangian but
corresponds to non-trivial non-local charges, and therefore is
expected to be related to the integrability of ${\cal N}$=4 SYM
\cite{{Mandal:2002fs},{Minahan:2002ve},{Bena:2003wd},Drummond:2009fd}.

\vskip 0.2cm

The dual conformal symmetry, via its anomaly Ward identity, can
uniquely fix the four and five-point amplitudes, which are the same
as the so-called BDS ansatz \cite{BDS} based on the explicit
perturbative calculations \cite{ABDK, BDS, 4p4l, 4p4l2}. However,
starting at six points, the amplitudes contain a function of cross
ratios which is not constrained by the dual conformal symmetry. This
function is called ``remainder function''\cite{AM2,
{two-loop-six-gluon}, {two-loop-six-Wilson-loop}}, which means that
it is an extra part that is not included in the BDS ansatz. The
studies of the remainder function have been done both at weak
coupling at two-loop level and in the strong coupling limit
\cite{Anastasiou:2009kna, Brandhuber:2009da, remainder-six,
remainder-six-simple, Alday:2009yn, Alday:2009dv, Alday:2010vh,
Alday:2010ku}.

\vskip 0.2cm

The problem of calculating amplitudes at strong coupling can be
solved by using the integrability of the classical world-sheet
theory \cite{Alday:2009yn, Alday:2009dv, Alday:2010vh}. The
essential point is that the equations of the system can be promoted
by including a spectral parameter $\zeta$ due to the integrability.
The cross ratios constructed from the solutions are therefore also
promoted to be functions of the spectral parameter. It then becomes
possible to obtain a set of functional relations between cross
ratios which can be organized as the so-called $Y$ system
\cite{Yang:1968rm,Zamolodchikov:1989cf}, where $Y$ functions are the
cross ratios. The boundary conditions can be nicely embedded via WKB
approximation \cite{Gaiotto:2008cd,Gaiotto:2009hg}, where the
dominant behavior of $Y$ functoin at large and small $\zeta$ is
given by the so-called WKB terms in which the mass parameters are
related to the shape of the polygon. Finally, the main non-trivial
part of the area can be given as the free energy of the $Y$ system.

\vskip 0.2cm

The general expression of amplitudes at strong coupling can be given
as \cite{Alday:2010vh, Yang:2010az}
\bea A  & = &  A_{\rm div} + A_{\rm BDS-like} + A_{\rm extra} +
A_{\rm periods} + A_{\rm free} ~. \eea
By comparing to the BDS ansatz which takes the form
\bea A^{\rm BDS-ansatz}  & = &  A_{\rm div} + A_{\rm BDS} ~, \eea
the reminder function at strong coupling can be given as
\bea R  & = &  (A_{\rm BDS-like} -A_{\rm BDS}) + A_{\rm extra} +
A_{\rm periods} + A_{\rm free} ~. \label{RF} \eea

The calculations are more tricky in the cases where the number of
gluons is a multiple of four $(n\!=\!4K)$. Such cases are special in
that a world-sheet coordinate transformation appearing in the
computation develops a non-trivial monodromy around infinity, which
makes the calculation more complicated, in particular for the
so-called cutoff part and periods part
\cite{Alday:2009yn,Yang:2010az}. In \cite{Yang:2010az}, it was shown
that the problem can be solved by introducing two extra equations
which involve non-adjacent kinematic invariants and also
$T$-functions. The terms in the cutoff part that depend on the
$T$-functions are defined as $A_{\rm extra}$, while the remaining
parts are defined as $A_{\rm BDS-like}$. A conjecture for the
periods part of $n\!=\!4K$ case was also made in \cite{Yang:2010az}
based on a generalization of the $AdS_3$ result \cite{Alday:2009yn}.
However, since the case of $AdS_5$ is much more complicated, it's
important to perform an honest calculation and check whether the
conjecture is correct or not. It is calculating the periods part for
the $n\!=\!4K$ case that motivated the study of the present paper.

\vskip 0.2cm

In this paper we study a special collinear limit of the scattering
amplitudes. This limit provides a strong restriction on the
amplitudes, in particular by which the periods part can be uniquely
fixed by the already calculated remaining parts of the amplitude.
The basic idea is very simple.

\vskip 0.2cm

Collinear limit (or more exactly the double-collinear limit) is the
limit where the momenta of two adjacent external particles become
parallel. The property of amplitudes in this limit provides strong
constraints for the results \cite{Bern:1995ix}. Particularly for the
amplitudes in ${\cal N}$=4 SYM, the BDS ansatz already accounts for
the collinear behavior of the full amplitudes \cite{BDS,
two-loop-six-gluon}, therefore the remainder function should have a
trivial behavior in the collinear limit, in the sense that a
$n$-point remainder function is directly reduced to a
$(n\!-\!1)$-point remainder function. It is interesting to consider
a series of collinear limits so that a $n$-side polygon is finally
reduced to a pentagon. Since pentagon has no non-trivial reminder
function (only a constant), the $n$-point reminder function is
reduced to a trivial constant in this special limit. This is the
limit on which we will focus in this paper.

\vskip 0.2cm

At strong coupling, this special limit has a nice picture in the
corresponding integrable system. The structure of minimal surface is
determined by a polynomial $P(z)$, where $z$ is the worldsheet
coordinate. For $n$-side polygon, the degree of the polynomial is
$(n-4)$. The shape of the polygon is determined by the coefficients
in the polynomial that are also related to the mass parameters of
the $Y$ system, which are defined as cycle integrals around two
zeros of the polynomial. Taking one mass parameter to be infinitely
large is equivalent to taking one of the zeros of the polynomial to
be infinity. The corresponding picture for the minimal surface is
that the a $n$-side polygon becomes a $(n\!-\!1)$-side polygon plus
a decoupled pentagon (which corresponds to the decoupled zero at
infinity). If all of the mass parameters are taken to be infinity,
then the $n$-side polygon is reduced to $(n\!-\!4)$ decoupled
pentagons. Therefore as already discussed, the remainder function in
this limit should be reduced to a trivial constant. This constraint
provides a way to calculate the periods part by using (\ref{RF}):
\bea A_{\rm periods} = - \Big[(A_{\rm BDS-like} - A_{\rm BDS}) +
A_{\rm extra} \Big] \Big|_{{\rm all~} m_s\rightarrow \infty} ~,
\label{periods-limit1} \eea
up to an constant. Notice that to obtain this relation we have used
two important facts. One is that the free energy part simply goes to
zero in this limit. The other one is that the periods part has the
structure of being quadratic in mass parameters. This is very
important so that we do not miss any information in the large mass
limit. This property is naturally expected from the definition of
the periods part as cycle integrals of the surface, while the
contributions which may go zero in the large mass limit are already
included in other parts, in particular for the $n\!=\!4K$ case in
the $A_{\rm extra}$ part which contains the information of
non-trivial non-compact cycles \cite{Alday:2009yn, Yang:2010az}.

\vskip 0.2cm

Although the idea is very simple and straightforward, in practice
there are a few technical issues to consider. One main issue is that
the (BDS$-$BDS-like) part on the right hand side of the formula is
expressed in terms of conventional cross ratios. To take the large
mass limit, one needs to write them as functions of mass parameters.
Generally the relations between cross ratios and mass parameters are
not simple, however, in the special limit that we consider these
relations turn out to be much simplified. Another complication is
that cross ratios are related to $Y$ functions in different phase
regions of the spectral parameter. When the phase shift is large,
the extra poles terms must be carefully included. This also happens
for the $A_{\rm extra}$ part which involves $T$ function at
different phase regions. We mention that there is also a shortcut
where there is no need to involve the discussion of pole
contributions by simply apply the functional relations, which was
explained in section \ref{section-5.1}.

\vskip 0.2cm

We study these issues in detail in this paper. The limit for
amplitudes up to twelve points are studied. For the known $n\neq 4K$
cases, we obtain the same results of periods part. This provides a
strong consistency check for the validity of our method. We then
applie it to calculate the periods part of eight and twelve-point
amplitudes for the first time.

\bigskip

The paper is organized as follows. In section \ref{Y-system} we give
a brief review of $Y$ system. In section \ref{crY}, we discuss the
relation between cross ratios and $Y$ functions. The pole
contribution due to large phase shift is considered in section
\ref{phasepole}. In section \ref{largemslimit} the large mass limit
of $Y$ functions and cross ratios is considered. In section
\ref{nneq4k}, we study the amplitudes of $n\neq 4K$ cases in this
limit. The eight and twelve-point cases are considered in section
\ref{neq4k}. Some useful formulae and results are collected in three
appendices.

\section{Basics of $Y$ system \label{Y-system}}

In this section we review some basic facts of $Y$ system. We mainly
provide the materials that are closely related to the problem
studied in this paper, which also set up our conventions. Readers
can find more details about the $Y$ system and its derivation for
amplitudes at strong coupling in \cite{Alday:2010vh, Yang:2010az}.

\vskip 0.2cm

We use the convention that
\bea f^\pm(\theta) \equiv f(\theta\pm i\pi/4) ~, \qquad
f^{[l]}(\theta) \equiv f(\theta+i l\pi/4) ~. \eea
where $\theta$ is the spectral parameter and can also be written as
$\zeta = e^\theta$.

We first introduce the  $T$ functions which are defined as
\bea T_{0,m} = \langle s_m s_{m+1} s_{m+2} s_{m+3} \rangle^{[-m-1]}
~, \qquad T_{4,m} = \langle s_{-2} s_{-1} s_0 s_1 \rangle^{[-m-1]}
~, \label{Tads5} \eea
\bea T_{1,m} = \langle s_{-2} s_{-1} s_0 s_{m+1} \rangle^{[-m]} ,
\quad T_{2,m} = \langle s_{-1} s_{0} s_{m+1} s_{m+2}
\rangle^{[-m-1]} , \quad T_{3,m} = \langle s_{-1} s_{m} s_{m+1}
s_{m+2} \rangle^{[-m]} ~, \nonumber\eea
where the contraction is defined as $\langle s_i s_{i+1} s_j s_{j+1}
\rangle \equiv \epsilon^{\alpha\beta\gamma\delta} s_{i,\alpha}
s_{i+1,\beta} s_{j,\gamma} s_{j+1,\delta}$.  These $s_i$ variables
are the smallest solutions of the flat equations which decay fastest
to the boundary\cite{Alday:2010vh}. Very interestingly they take the
same form of the momentum twistor variables, which were first
introduced at weak coupling \cite{Hodges:2009hk} (see also
\cite{Mason:2009qx}), and are related to ordinary Lorentz variables
as
\bea x_{ij}^2 = { X_i\cdot X_j \over X_i^+ X_j^+ } ~, \qquad
X_i\cdot X_j = \langle s_i s_{i+1} s_j s_{j+1} \rangle ~, \qquad
X_i^{\alpha\beta} \sim s_i^\alpha \wedge s_{i+1}^\beta ~.
\label{xij2} \eea

The $Y$ functions can be constructed from $T$ functions as
\bea Y_{a,m} = {T_{a,m-1} T_{a,m+1} \over T_{a-1,m} T_{a+1,m}} ~.
\label{YfromT} \eea
They are related to cross ratios as we will see later.

We can impose the normalization conditions
\bea \langle s_i s_{i+1} s_{i+2} s_{i+3} \rangle = 1 ~,
\label{s1234} \eea
with which we have
\bea T_{0,m} = T_{4,m} = T_{a,0} = 1 ~. \eea
There are also important shifting relations provided by the $Z_4$
symmetry of the corresponding $SU(4)$ Hitchin system
\cite{Alday:2010vh} as
\bea && \langle s_{j-1} s_j s_{k-1} s_k \rangle^{[2]} = \langle s_j
s_{j+1} s_k s_{k+1} \rangle ~, \label{z4shift} \\ && \langle s_{j-2}
s_{j-1} s_j s_k \rangle^{[2]} = \langle s_j s_k s_{k+1} s_{k+2}
\rangle ~,
\\ && \langle s_j s_{k-2} s_{k-1} s_k \rangle^{[2]} = \langle s_j
s_{j+1} s_{j+2} s_k \rangle ~. \eea

The $T$ functions satisfy the Hirota equation
\bea T_{a,s}^+ T_{4-a,s}^- = T_{4-a,s+1} T_{a,s-1} + T_{a+1,s}
T_{a-1,s} ~, \label{Hirota} \eea
where $a=1,2,3$, and for $n$-point, $s=1,2,...,n-5$. This also
implies the following functional relations for $Y$ functions
\bea { Y_{2,s}^- Y_{2,s}^+ \over Y_{1,s} Y_{3,s} } & = & {
(1+Y_{2,s+1}) (1+Y_{2,s-1}) \over (1+Y_{1,s})(1+Y_{3,s})} ~, \label{Y-functional1} \\
{ Y_{3,s}^- Y_{1,s}^+ \over Y_{2,s} } & = & {
(1+Y_{3,s+1}) (1+Y_{1,s-1}) \over 1+Y_{2,s}} ~, \\
{ Y_{1,s}^- Y_{3,s}^+ \over Y_{2,s} } & = & { (1+Y_{1,s+1})
(1+Y_{3,s-1}) \over 1+Y_{2,s}} ~. \label{Y-functional3} \eea
These functional relations can be written in equivalent but
practically more useful integral forms as
\bea \log Y_{2,s} & = & - \sqrt{2} m_s \cosh\theta - K_2 \star
\alpha_s - K_1 \star \beta_s ~,  \label{logY2s} \\ \log Y_{1,s} & =
& - m_s \cosh\theta - C_s - {1\over2} K_2 \star \beta_s - K_1 \star
\alpha_s
- {1\over2} K_3 \star \gamma_s ~, \\
\log Y_{3,s} & = & - m_s \cosh\theta + C_s - {1\over2} K_2 \star
\beta_s - K_1 \star \alpha_s + {1\over2} K_3 \star \gamma_s ~, \eea
where the terms including mass parameters $m_s$ and $C_s$ are the
WKB terms that are related to the boundary conditions. The other
functions are defined as
\bea && \alpha_s \equiv \log {(1+Y_{1,s})(1+Y_{3,s}) \over
(1+Y_{2,s-1})(1+Y_{2,s+1}) } ~, \qquad \gamma_s \equiv \log
{(1+Y_{1,s-1})(1+Y_{3,s+1}) \over (1+Y_{1,s+1})(1+Y_{3,s-1}) } ~, \nonumber \\
&& \beta_s \equiv \log { (1+Y_{2,s})^2 \over
(1+Y_{1,s-1})(1+Y_{1,s+1})(1+Y_{3,s-1})(1+Y_{3,s+1})} ~.
\label{abgamma} \eea
and the kernels are
\bea K_1(\theta) \equiv {1 \over 2 \pi} {1\over\cosh \theta} ~,
\qquad K_2(\theta) \equiv {\sqrt{2} \over \pi}
{\cosh\theta\over\cosh 2\theta} ~, \qquad K_3(\theta) \equiv {i
\over \pi} {\tanh 2\theta} ~. \label{kernel} \eea
Notice the ``$\star $'' operation can be written explicitly as
\bea K_i \star f_s = \int_{-\infty}^{+\infty} d\theta' ~
K_i(\theta-\theta') f_s(\theta') ~. \eea
In this form we take the phase $m_s$ to be real positive, and the
range for the phase of $\zeta$ is $\phi = {\rm Im}(\theta) \in
(-\pi/4, \pi/4)$. In other regions, the above integral equations
need to be modified. We will discuss this point in more detail
later.

The integral form of $T$ functions can also be given as
\cite{Yang:2010az}
\bea \log T_{2,1} & = & K_2 \star \log (1+ Y_{2,1}) +K_1 \star \log
(1+Y_{1,1})(1+Y_{3,1}) ~, \label{T21} \\ \log T_{1,1} & = &
{1\over2} K_2 \star \log (1+ Y_{1,1}) (1+ Y_{3,1}) + K_1 \star \log
(1+Y_{2,1}) + {1\over2} K_3 \star
\log {(1+ Y_{1,1}) \over (1+ Y_{3,1})} ~, \label{T11} \\
\log T_{3,1} & = & {1\over2} K_2 \star \log (1+ Y_{1,1}) (1+
Y_{3,1}) + K_1 \star \log (1+Y_{2,1}) - {1\over2} K_3 \star \log
{(1+ Y_{1,1}) \over (1+ Y_{3,1})} ~. \label{T31} \eea
All other $T$ functions can be obtained by using these three $T$
functions and $Y$ functions. For example, by using the relation
$T_{2,2} = T_{1,1} T_{3,1} Y_{2,1}$, we can obtain  $T_{2,2}$ as
\bea \log T_{2,2} & = & - \sqrt{2} m_1 \cosh(\theta) + K_2 \star
\log {(1+Y_{2,2}) } + K_1 \star \log { (1+Y_{1,2})(1+Y_{3,2})} ~.
\label{T22} \eea

We emphasize that a special normalization of $T$ functions is
chosen, for example one can notice that there is no WKB term in
$T_{2,1}$. This freedom of choosing normalization comes from the
gauge redundancy of the Hirota equation for $T$ functions
\cite{Alday:2010vh}. But the full amplitudes are of course
independent of such choice. Since our method is based on the
consistency condition of the full amplitudes, we do not need to
worry about this issue here.

\section{Cross ratios and $Y$ functions \label{crY}}

As mentioned before, the amplitudes at strong coupling involve
parameters $m_s$, which are in general related to cross ratios
though complicated integral equations of $Y$ functions. To study the
collinear limit of amplitudes, we first need to study the relation
between cross ratios and $Y$ functions. We will also see that one
can obtain a one parameter family of cross ratios which will provide
further restriction on the remainder functions.

\subsection{From $Y$ functions to cross ratios \label{Y-cr}}

The traditional cross ratios which involve consecutive cusps can be
defined as
\bea u_{ij} \equiv { x_{i,j+1}^2 x_{i+1,j}^2 \over x_{ij}^2
x_{i+1,j+1}^2} = {\langle s_i s_{i+1} s_{j+1} s_{j+2} \rangle
\langle s_{i+1} s_{i+2} s_j s_{j+1} \rangle \over \langle s_i
s_{i+1} s_{j} s_{j+1} \rangle \langle s_{i+1} s_{i+2} s_{j+1}
s_{j+2} \rangle } ~, \eea
where we have written the cross ratios in terms of the small
solutions by using (\ref{xij2}). A generic cross ratio involving
non-consecutive cusps can be constructed by multiplying together
consecutive ones, for example
\bea { x_{i,j+2}^2 x_{i+2,j}^2 \over x_{ij}^2 x_{i+2,j+2}^2} =
u_{ij} u_{i,j+1} u_{i+1,j} u_{i+1,j+1} ~. \eea
The cross ratios are not all independent. For $n$-point where
$n\!>\!5$, there are only $3(n-5)$ independent cross ratios, due to
the constraints of on-shell condition and Gram determinant
constraints \cite{Drummond3, Anastasiou:2009kna}.

The consecutive-cusp cross ratios can be constructed via $Y_{2,m}$
by using (\ref{Tads5})-(\ref{YfromT}) as
\bea {Y_{2,m}\over 1+ Y_{2,m}} =
{ \langle s_{-1} s_{0} s_{m+2} s_{m+3} \rangle^{[-m-2]} \langle
s_{-1} s_{0} s_{m} s_{m+1} \rangle^{[-m]} \over \langle s_{-1} s_{0}
s_{m+1} s_{m+2} \rangle^{[-m-2]} \langle s_{-1} s_{0} s_{m+1}
s_{m+2} \rangle^{[-m]} } ~. \eea
When $m=2k-2$, we have
\bea {Y_{2,2k-2} \over 1+ Y_{2,2k-2}} =  { \langle s_{-k-1} s_{-k}
s_k s_{k+1} \rangle \langle s_{-k} s_{-k+1} s_{k-1} s_k \rangle
\over \langle s_{-k-1} s_{-k} s_{k-1} s_k \rangle \langle s_{-k}
s_{-k+1} s_{k} s_{k+1} \rangle } = {x^2_{-k-1,k} x^2_{-k,k-1} \over
x^2_{-k-1,k-1} x^2_{-k,k} } ~. \eea
When $m\!=\!2k\!-\!1$, we have
\bea \left( {Y_{2,2k-1} \over 1+ Y_{2,2k-1}} \right)^+ =  { \langle
s_{-k-1} s_{-k} s_{k+1} s_{k+2} \rangle \langle s_{-k} s_{-k+1}
s_{k} s_{k+1} \rangle \over \langle s_{-k-1} s_{-k} s_{k} s_{k+1}
\rangle \langle s_{-k} s_{-k+1} s_{k+1} s_{k+2} \rangle }
 =  { x^2_{-k-1,k+1} x^2_{-k,k} \over
x^2_{-k-1,k} x^2_{-k,k+1} } ~. \eea
Using the shifting relation (\ref{z4shift}), we can write all
consecutive cross ratios in terms of $Y$ functions.

As an explicit example, for the simplest six-point case, there are
three independent cross ratios
\bea u_{14} = {x_{15}^2 x_{24}^2 \over x_{14}^2 x_{25}^2} ~, \qquad
u_{25} = {x_{26}^2 x_{35}^2 \over x_{25}^2 x_{36}^2} ~. \qquad
u_{36} = {x_{13}^2 x_{46}^2 \over x_{36}^2 x_{14}^2} ~. \eea
We can express them in terms of $Y$ functions as
\bea u_{36} = {Y_{2,1}^-\over 1+Y_{2,1}^-} ~, \qquad u_{14} =
{Y^+_{2,1}\over 1+Y^+_{2,1}} ~, \qquad u_{25} = {Y^{[3]}_{2,1}\over
1+Y^{[3]}_{2,1}} ~. \label{u36} \eea
The relations between $Y$ functions and the cross ratios that are
commonly used in weak coupling calculation for higher-point cases
are given explicitly in appendix \ref{Y-cr}.

\subsection{One parameter family of cross ratios}

$Y$ functions (and therefore also cross ratios) are functions of
spectral parameter. The physical cross ratios of the original
problem can be recovered by taking $\zeta = e^\theta = 1$. A very
interesting fact is that one can actually obtain a one parameter
family of physical cross ratios by taking $\zeta$ to be a pure
phase, or equivalently $\theta$ to be a pure imaginary number
$i\phi$ \cite{Alday:2010vh}. We can write this one family of cross
ratios by including the $\phi$-dependence as
\bea u_{36}(i\phi) = {Y_{2,1}^-(i\phi)\over 1+Y_{2,1}^-(i\phi)} ~,
\qquad u_{14}(i\phi) = {Y^+_{2,1}(i\phi) \over 1+Y^+_{2,1}(i\phi)}
~, \qquad u_{25}(i\phi) = {Y^{[3]}_{2,1}(i\phi)\over
1+Y^{[3]}_{2,1}(i\phi)} ~. \label{six-cr} \eea

An important fact is that the free energy and periods parts are
independent of the phase $\phi$, and therefore take same values for
the whole family of cross ratios. This is due to the integrability
of the world-sheet theory\cite{Alday:2010vh}\footnote{It is very
interesting to see if there is any similar invariance at weak
coupling, which may provide a connection between weak coupling
Yangian symmetry and the integrability at strong coupling}.
This invariance provides a further constraint for the amplitudes,
which is important for the limit that we consider in this paper.
Although $A_{\rm free}$ and $A_{\rm periods}$ are independent of the
phase, the other parts of remainder function do depend. Therefore,
the remainder function also depends on $\phi$ as
\bea R(\phi)  & = &  (A_{\rm BDS-like} -A_{\rm BDS})(\phi) + A_{\rm
extra}(\phi) + A_{\rm periods} + A_{\rm free} ~.  \eea
On the other hand, the limit we consider which gives rise to
(\ref{periods-limit1}) is still true, therefore we have that
\bea A_{\rm periods} = - \Big[ (A_{\rm BDS-like} -A_{\rm BDS})(\phi)
+ A_{\rm extra}(\phi)\Big] \Big|_{{\rm all~} m_s\rightarrow \infty}
~. \eea
This provides a stronger version of (\ref{periods-limit1}). Since
each part on the right hand side depends on the phase separately,
there must be non-trivial cancellation, so that the finally result
is independent of $\phi$ as required by the equality. We will see
that this is indeed true. This restriction provides another
important consistency check for our method.

\section{Phase shift and pole terms \label{phasepole}}

To obtain cross ratio, we need to consider $Y$ functions with large
phase shift
\bea Y_{2,s}^{[r]}(\theta) = Y_{2,s}(\theta + i {r \pi / 4} ) ~.
\eea
This will require a modification of the integral equations. The
reason is that residue contributions should be included in the
integral equations when we cross the pole lines of kernel functions.
Mathematically, this is similar to the wall crossing phenomenon in
the study of supersymmetric field theory \cite{Gaiotto:2008cd,
Gaiotto:2009hg}.

In general there are two cases that one needs to consider pole
contribution. One is the argument $\theta$ of $Y$ functions is
fixed, but the phase of mass parameters changes. The other case is
what we have encountered above, where the mass parameters are fixed,
but the argument $\theta$ (or more exactly its imaginary part) of
$Y$ functions varies. The mathematical reason for both cases is the
same, but the physical pictures are different. We are mainly
interested in the latter case. We will take the mass parameters as
real positive numbers\footnote{Notice that for simplicity, we take
$m_s$ to be real positive number. One can of course include a phase
$\varphi_s$ for each $m_s$. Then one may need to modify the
equations also according to the range of $\varphi_s$
\cite{Alday:2010vh}. The discussion will be more involved but
straightforward.}, then the integral equations for $Y$ functions
will preserve their original form as long as $\phi={\rm Im} (\theta)
\in (-\pi/4, \pi/4)$
\bea \log Y_{2,s}(\theta) & = & - \sqrt{2} m_s \cosh(\theta) - K_2
\star \alpha_s - K_1 \star \beta_s ~, \\ \log Y_{1,s}(\theta) & = &
- m_s \cosh(\theta) - C_s - {1\over2} K_2 \star \beta_s - K_1 \star
\alpha_s
- {1\over2} K_3 \star \gamma_s ~, \\
\log Y_{3,s}(\theta) & = & - m_s \cosh(\theta) + C_s - {1\over2} K_2
\star \beta_s - K_1 \star \alpha_s + {1\over2} K_3 \star \gamma_s ~.
\eea
Beyond that region, one needs to modify the equations by including
extra pole contributions while $\theta$ crosses the lines of $\pm
i\pi/4, \pm i\pi/2, \cdots$ of $\zeta$ plane. Explicitly, the
positions of the poles $\theta_p$ of each kernel are
\bea K_1 = {1 \over 2 \pi} {1\over\cosh \theta} ~, &&  \qquad
\theta_p = i{(1+2l)\pi\over2} ~, \\
K_2 = {\sqrt{2} \over \pi} {\cosh\theta\over\cosh 2\theta} ~,
&& \qquad \theta_p = i{(1+2l)\pi\over4} ~, \\
K_3 = {i \over \pi} {\tanh 2\theta} ~, && \qquad \theta_p =
i{(1+2l)\pi\over4} ~, \qquad \textrm{$l$ is integer}~.  \eea

As an explicit example, we consider $Y_{1,s}(\theta)$ in the case
where ${\rm Im}(\theta) \in (3\pi/4, \pi)$. It crosses three pole
lines. While crossing the line ${\rm Im} (\theta) = \pi/4$, we have
pole contributions from kernels $K_2$ and $K_3$. When crossing the
line ${\rm Im} (\theta) = \pi/2$, we have pole contribution from
kernels $K_1$. And while crossing the line ${\rm Im} (\theta) =
3\pi/4$, we have pole contributions from kernels $K_2$ and $K_3$
again. In total, for $Y_{1,s}(\theta)$ with argument ${\rm
Im}(\theta) \in (3\pi/4, \pi)$ we have
\bea && \log Y_{1,s}(\theta) = - m_s \cosh(\theta) - C_s - {1\over2}
K_2 \star \beta_s - K_1 \star \alpha_s - {1\over2} K_3 \star
\gamma_s \nonumber\\ && \hskip 0.5cm -
{1\over2}\beta_s(\theta-i\pi/4) + {1\over2}\gamma_s(\theta-i\pi/4) -
\alpha_s(\theta-i\pi/2)  - {1\over2}\beta_s(\theta-i3\pi/4) +
{1\over2}\gamma_s(\theta-i3\pi/4) ~. \qquad \eea
Notice that one should also be careful about the signs of residues.

It is convenient to use another notation when we consider cross
ratios. For example, for six-point we want to calculate
\bea Y_{2,1}^-(i\phi) = Y_{2,1}(i\phi-i\pi/4) ~, \quad
Y_{2,1}^+(i\phi) = Y_{2,1}(i\phi + i\pi/4) ~, \quad
Y_{2,1}^{[3]}(i\phi) = Y_{2,1}(i\phi+i3\pi/4) ~. \eea
We choose $\phi \in (0, \pi/4)$. We can see that there is no pole
contribution for $Y_{2,1}^-(i\phi)$, but there are pole
contributions for the other two functions. Similar to the example
above, we can obtain that
\bea \log Y_{2,1}^-(i\phi) &=& - \sqrt{2} m_1 \cos(\phi-\pi/4) - K_2
\star \alpha_1 - K_1 \star \beta_1 ~, \label{Y21m1} \\ \log
Y_{2,1}^+(i\phi) &=& - \sqrt{2} m_1 \cos(\phi+\pi/4) - K_2 \star
\alpha_1 - K_1 \star \beta_1 - \alpha_1(i\phi) ~, \\ \log
Y_{2,1}^{[3]}(i\phi) &=& - \sqrt{2} m_1 \cos(\phi+3\pi/4) - K_2
\star \alpha_1 - K_1 \star \beta_1 - \alpha_1^{[2]}(i\phi) -
\beta_1^{[1]}(i\phi) - \alpha_1(i\phi) ~. ~~~~~ \label{Y21-3} \eea
We can then define the cross ratios as in
(\ref{six-cr})\footnote{One may notice that the integral equations
for cross ratios looks a little different from that in
\cite{Alday:2009dv} (see also \cite{Hatsuda:2010vr}). They are
equivalent but written in different presentations, and are related
by the functional relations. One can also explicitly solve the the
integral equations numerically and find they give the same cross
ratios.}. In appendix \ref{Y-phase}, we provide equations for more
complicated cases which we will encounter when we calculate higher
points amplitudes. Since we need to calculate $T$ functions when
there is $A_{\rm extra}$ part, the equations for $T_{2,1}, T_{2,2}$
with large phase shift are also given.

We should mention that these integral equations are correct no
matter whether we take the large mass limit or not. Therefore they
can be used for more general studies. Numerically they can be solved
very efficiently at least when the number of gluons is not very
large. There are two very useful checks for the equations and
results. One is that in the small $m_s$ limit, the $Y$ functions
should converge to the same value which is independent of $\phi$.
This corresponds to the regular polygon limit, where all cross
ratios take the same value. The other more non-trivial check is the
continuous condition that, $Y$ functions themselves must be
continuous functions for all phase, despite that the expressions of
$Y$ functions in different regions of phase look different due to
extra pole terms. In other words, one needs to check that there is
no discontinuity when the pole lines are crossed. We have done these
checks for the equations we give.

\section{Large mass limit of $Y$ functions\label{largemslimit}}

When all $m_s$ are taken to be infinitely large, the relations
between cross ratios and mass parameters become much simpler. The
main reason is that the complicated integral terms always vanish in
the limit, as
\bea K \star \log(1+ Y_{a,s}) = \int_{-\infty}^{+\infty} d\theta ~ K
\log (1+ e^{- m_s \cosh\theta + \cdots}) & \rightarrow & 0 ~.
\label{limit-integral} \eea
Notice that for simplicity we take $m_s$ to be real positive.
Therefore, the $Y$ functions are dominated by the WKB terms in the
limit. The main complication comes from that the pole terms may also
have contribution in the limit.

We consider six-point case explicitly. We first consider $u_{36}$ in
(\ref{u36}). In the large $m_1$ limit, we can neglect the integral
terms in (\ref{Y21m1}) and have
\bea \log Y_{2,1}^-(i\phi) ~~ \rightarrow ~~ - \sqrt{2} m_1
\cos(\phi-\pi/4) ~.  \eea
Then we have
\bea u_{36}(i\phi) = {Y_{2,1}^{[-1]}(i\phi)\over
1+Y_{2,1}^{[-1]}(i\phi)} \rightarrow { e^{- \sqrt{2} m_1
\cos(\phi-\pi/4)} \over 1+ e^{- \sqrt{2} m_1 \cos(\phi-\pi/4)} }
\rightarrow e^{- \sqrt{2} m_1 \cos(\phi-\pi/4)} ~. \eea
Notice that the sign of $\log Y$ in the limit makes the results
totally different. Here it is negative. If it is positive, then $Y
\rightarrow + \infty$ and the corresponding cross ratio will simply
be one in the limit.

For $u_{14}$ we need to consider $Y_{2,1}^+(i\phi)$, where there is
a pole term
\bea \alpha_1(i\phi) &=& \log (1+Y_{1,1}(i\phi)) +
\log(1+Y_{3,1}(i\phi)) \nonumber \\ &\rightarrow & \log (1+ e^{- m_1
\cos(\phi) + C_1} ) + \log(1+ e^{- m_1 \cos(\phi)- C_1 } )  ~. \eea
In the second line we take the limit and neglect the integral terms
in $Y_{1,1}$ and $Y_{3,1}$. Comparing to WKB terms in $Y_{2,1}^+$,
this pole contribution is exponentially suppressed and can be
neglected, therefore we have
\bea \log Y_{2,1}^+(i\phi) \rightarrow - \sqrt{2} m_1
\cos(\phi+\pi/4) ~, \eea
and the cross ratio is
\bea u_{14}(i\phi) \rightarrow e^{- \sqrt{2} m_1 \cos(\phi+\pi/4)}
~. \eea

Similarly we consider $u_{25}$. We can see from (\ref{abgamma}) that
the pole terms are always a summation of functions in the form of
$\log (1+ Y)$, therefore when $Y \rightarrow e^{-\infty}$ the
contribution can always be neglected. But when $Y \rightarrow
e^{+\infty}$, there is non-trivial contribution, which turns out to
be the case for the pole term $\alpha_1^{[2]}(i\phi)$ in
$Y_{2,1}^{[3]}(i\phi)$. We have
\bea \alpha_1^{[2]}(i\phi) &=& \log
{(1+Y_{1,1}^{[2]}(i\phi)) + \log(1+Y_{3,1}^{[2]}(i\phi))  } \nonumber\\
&\rightarrow& \log {(1 + e^{- m_1 \cos(\phi+\pi/2) + C_1} ) +
\log(1+ e^{- m_1 \cos(\phi+\pi/2) - C_1 } )  }  \nonumber\\
&\rightarrow& ( m_1 \sin(\phi) + C_1 ) + ( m_1 \sin(\phi) - C_1  )   \nonumber\\
&\rightarrow& 2 m_1 \sin(\phi)  ~. \label{pole-alpha}\eea
Notice the parameters $C_s$ always cancel with each other in the
limit, since $Y_{1,s}$ and $Y_{3,s}$ always appear in pairs. The
other pole terms in (\ref{Y21-3}) are exponential suppressed,
therefore in total we have
\bea \log Y_{2,1}^{[3]}(i\phi) &\rightarrow& \sqrt{2} m_1
\sin(\phi+\pi/4) - 2 m_1 \sin(\phi) ~. \eea
We can see that $Y_{2,1}^{[3]} \rightarrow +\infty$ for
$\phi\in(0,\pi/4)$, therefore $u_{25}$ is simply one in the limit.

To summarize we have the cross ratios in the large mass limit as
\bea \{ u_{36}(i\phi) , ~  u_{14}(i\phi) , ~  u_{25}(i\phi) \} &
\rightarrow & \{ e^{-\sqrt{2} m_1 \cos(\phi-\pi/4)} , ~ e^{-\sqrt{2}
m_1 \cos(\phi+\pi/4)} , ~ 1 \} ~. \eea
The results including the cross ratios of several higher-point cases
and $T$ functions will be given in the following sections. They can
be derived in the same way.

\subsection{Another way bypassing the pole contributions \label{section-5.1}}

There is another way to obtain the large mass limit of the cross
ratios in which one does not need to consider the pole
contributions%
\footnote{We would like to thank the anonymous referee for this
suggestion.}. The idea is that one can always use the functional
relations (\ref{Y-functional1})-(\ref{Y-functional3}) to bring
$Y_{a,s}^{[r]}$ to a combination of $Y_{a,s}^{[r-1]}$ and
$Y_{a,s}^{[r-2]}$, or to a combination of $Y_{a,s}^{[r+1]}$ and
$Y_{a,s}^{[r+2]}$. Thus it is always possible to express all cross
ratios in terms of only $Y_{a,s}^{[0]}$ and $Y_{a,s}^{[-1]}$ for
which the phases are in the region $\phi \in (-\pi/4, \pi/4)$. Then
one can simply apply the relation
\bea \log Y_{2,s}(i\phi) &\simeq& - \sqrt{2} m_s \cosh(\phi) ~, \nonumber \\
~ \log Y_{1,s}(i\phi) &\simeq& - m_s \cosh(\phi) - C_s  ~, \\ \log
Y_{3,s}(i\phi) &\simeq& - m_s \cosh(\phi) + C_s \nonumber \eea
to obtain the large mass limit of the cross ratios.

Using Hirota equation (\ref{Hirota}), one can similarly calculate
the $T$ functions which appear in $A_{\rm extra}$ part in the same
way.

\section{Consistency check for the known results \label{nneq4k}}

We first apply out method to the $n\neq 4K$ case, where there is no
$A_{\rm extra}$ part. The periods can be calculated as
\bea A_{\rm periods} = - \Big[ (A_{\rm BDS-like} - A_{\rm
BDS})(\phi) \Big] \Big|_{{\rm all~} m_s\rightarrow \infty}  ~ . \eea
We will reproduce all known results for up to eleven points. We
summarize the results of amplitudes which may be used here in
appendix \ref{results}.

\subsection{Six-point}

The results of periods part and the difference between BDS-like and
BDS part are known \cite{Alday:2009dv}
\bea && A_{\rm periods}^{n=6} = {|m_1|^2 \over 4} ~, \label{6periods} \\
&& A_{\rm BDS-like}^{n=6} - A_{\rm BDS}^{n=6} = -\sum_{i=1}^3 \left(
{1\over8}\log^2 u_{i,i+3} + {1\over4} Li_2(1-u_{i,i+3})\right) ~.
\label{six-bdslikebds} \eea
The cross ratios in the large mass limit have been obtained above
\bea \{ u_{36}(i\phi) , ~  u_{14}(i\phi) , ~  u_{25}(i\phi) \} &
\rightarrow & \{ e^{-\sqrt{2} m_1 \cos(\phi-\pi/4)} , ~ e^{-\sqrt{2}
m_1 \cos(\phi+\pi/4)} , ~ 1 \} ~. \eea
Applying this limit in (\ref{six-bdslikebds}) we obtain that
\bea (A_{\rm BDS-like}^{n=6} - A_{\rm BDS}^{n=6})(\phi)
&\rightarrow& - \left( {1\over8} 2m_1^2[\cos^2(\phi-\pi/4) +
\cos^2(\phi+\pi/4)] + {1\over4} Li_2(1)\right) \nonumber\\ &=&
-{1\over4} m_1^2 + {\rm constant} ~, \eea
which indeed cancels with $A_{\rm periods}$ up to a constant.
Although we have choose $m_s$ to be real positive number for
simplicity, this is enough to reproduce the more general expression
(\ref{6periods}) by knowing that the periods part is real. One can
of course also do an honest calculation, for example consider to
include a small phase factor for $m_s$ (so that the equations do not
change) and find that the result is indeed correct.

We can see that in the limit, although cross ratios still depends on
the phase $\phi$, the difference between the BDS-like and BDS part
is independent of $\phi$. We emphasize that this independence is not
true in general, it only happens in the limit, which is required to
cancel the periods part. As we will see in the $n=4K$ case, such
cancellation will become more non-trivial due to the existence of
extra part.

\subsection{Seven-point}

We have the result
\bea   A_{\rm periods}^{n=7} = {|m_1|^2 + |m_2|^2 \over 2} + {m_1
\bar m_2 + \bar m_1 m_2 \over 2 \sqrt{2}} ~, \label{7periods} \hskip 8.6cm \\
A_{\rm BDS-like}^{n=7} - A_{\rm BDS}^{n=7} = -{1\over 4}\sum_{i=1}^7
\left( \log^2 u_{i,i+3} + Li_2(1-u_{i,i+3}) -{1\over2} \log
u_{i,i+3} \log {u_{i+2,i+5} u_{i+1,i+5} \over u_{i+3,i+6} u_{i,i+4}}
\right) \label{7bdsdif} ~. \eea
The cross ratios in the large mass limit can be obtained as
\bea && \{ u_{14}(i\phi) , ~  u_{25}(i\phi) , ~ u_{36}(i\phi) , ~
u_{47}(i\phi) , ~ u_{15}(i\phi) , ~
 u_{26}(i\phi) , ~  u_{37}(i\phi) \} \rightarrow \nonumber\\ && \hskip 1cm
\{ e^{-\sqrt{2} m_2\cos(\phi)} , ~1, ~ 1, ~ e^{-\sqrt{2}
m_1\cos(\phi-\pi/4)}, ~ e^{-\sqrt{2} m_1\cos(\phi+\pi/4)} , ~1 , ~
e^{-\sqrt{2} m_2\sin(\phi)} \} ~. \qquad \eea
Applying this limit in (\ref{7bdsdif}) we find that
\bea  (A_{\rm BDS-like}^{n=7} - A_{\rm BDS}^{n=7})(\phi) \rightarrow
-{ m_1^2 + m_2^2 \over 2} - { m_1 m_2 \over \sqrt{2}} + {\rm
constant} ~, \eea
which is independent of the $\phi$, and consistent with the known
result of periods part (\ref{7periods}).

\subsection{Nine-point}

We have the results that
\bea && A_{\rm periods}^{n=9} = -{1\over2} (|m_2|^2 + |m_3|^2 + m_1
\bar m_3 + m_3 \bar m_1 + m_2 \bar m_4 + m_4 \bar m_2 )  \label{9periods} \\
&& \qquad\qquad - {\sqrt{2} \over 4} (m_1 \bar m_2 + m_2 \bar m_1 +
m_1 \bar m_4 + m_4 \bar m_1 + m_3 \bar m_4 + m_4 \bar m_3 + 2 m_2
\bar m_3 + 2 m_3 \bar m_2)
\nonumber ~, \\ &&  A_{\rm BDS-like}^{n=9} - A_{\rm BDS}^{n=9} =
{1\over 4}\sum_{i=1}^9 \left[ \log u_{i,i+3} \log \left( {u_{i,i+4}
u_{i+4,i+7} \over u_{i+1,i+5} u_{i+2,i+5} u_{i+3,i+6} u_{i+2,i+6}^2}
\right) \right.
\\ && \hskip 1cm \left. + \log u_{i,i+4} \log \left(
{u_{i,i+4} u_{i+1,i+4} u_{i+4,i+8}^2 u_{i+5,i+8}^2 \over u_{i+2,i+5}
u_{i+2,i+6}^3 u_{i+3,i+6}^2 u_{i+3,i+7}^2} \right) - Li_2 \left( 1-
u_{i,i+3} \right) - Li_2 \left( 1- u_{i,i+4} \right) \right] ~.
\nonumber \eea
The cross ratios in the large mass limit can be obtained as
\bea && \{ u_{14}(i\phi) , ~  u_{25}(i\phi) , ~ u_{36}(i\phi) , ~
u_{47}(i\phi) , ~ u_{58}(i\phi) , ~
 u_{69}(i\phi) , ~  u_{17}(i\phi), ~  u_{28}(i\phi), ~  u_{39}(i\phi) \}  \rightarrow
 \nonumber\\ && \qquad \{ e^{-\sqrt{2} m_4\sin(\phi)}, ~ e^{-\sqrt{2} m_4\cos(\phi)}, ~ 1
,~ 1 , ~1, ~ e^{-\sqrt{2} m_1 \cos(\phi-\pi/4)}, ~ e^{-\sqrt{2} m_1 \cos(\phi+\pi/4)} , ~1, ~ 1 \} ~, \\
&& \{ u_{15}(i\phi) , ~ u_{26}(i\phi) , ~ u_{37}(i\phi) , ~
u_{48}(i\phi), ~ u_{59}(i\phi) , ~
u_{16}(i\phi) , ~ u_{27}(i\phi) , ~ u_{38}(i\phi) ,~ u_{49}(i\phi) \}  \rightarrow  \nonumber\\
&& \qquad \{ e^{-\sqrt{2} m_3 \cos(\phi-\pi/4)}, ~ e^{-\sqrt{2} m_3
\cos(\phi+\pi/4)} , ~1, ~1, ~e^{-\sqrt{2} m_2\sin(\phi)}, ~
e^{-\sqrt{2} m_2\cos(\phi)}, ~ 1 ,~ 1 , ~1 \} ~. \qquad\qquad  \eea
One can find in this limit that
\bea && (A_{\rm BDS-like}^{n=9} - A_{\rm BDS}^{n=9})(\phi)
\rightarrow \nonumber\\ && \quad {1\over2} (m_2^2 + m_3^2 + 2 m_1
m_3 + 2 m_2 m_4) + {\sqrt{2} \over 2} ( m_1 m_2  + m_1  m_4 + m_3
m_4 +2 m_2 m_3 ) + {\rm constant} ~, \qquad \eea
which is independent of the $\phi$, and consistent with the result
of periods part (\ref{9periods}).

\vskip 0.3cm

We also calculate the ten and eleven-point cases and also reproduce
the periods part given in \cite{Alday:2010vh}. Since these checks
are very non-trivial, it provides a strong check for the validity of
our method. We will consider $n=4K$ cases in next section.

\section{Fix the periods part for $n=4K$ cases \label{neq4k}}

For $n=4K$ case, we can calculate the periods parts by using
\bea A_{\rm periods} = - \Big[ (A_{\rm BDS-like} -A_{\rm BDS})(\phi)
+ A_{\rm extra}(\phi)\Big] \Big|_{{\rm all~} m_s\rightarrow \infty}
~. \label{constraint} \eea

For eight point, the cutoff part of the amplitudes was calculated in
\cite{Yang:2010az}. We can obtain that
\bea A_{\rm BDS-like}^{n=8} - A_{\rm BDS}^{n=8} &=& - {1\over8}
\log^2 \left( {u_{14} u_{25} u_{58} u_{16} \over u_{36} u_{47}
u_{27}} \right) - {1\over4} \log \left( u_{15} u_{16} u_{25} u_{26}
\right) \log \left( u_{26} u_{27} u_{36} u_{37} \right) \nonumber\\
&& - {1\over2} \log \left( {u_{15} \over u_{37}} \right) \log \left(
{u_{14} u_{25} u_{58} u_{16} u_{15} \over u_{36} u_{47} u_{27}
u_{37}} \right)
- {1\over8} (\Delta_x - \Delta_y)^2 \nonumber\\
&& + {1\over4} \Delta_x \log \left( {u_{14} u_{15}^2 u_{16} u_{25}
u_{58} \over u_{27} u_{36} u_{37}^2 u_{47}} \right) + {1\over4}
\Delta_y \log \left( {u_{26} u_{27} u_{36} u_{37}^2 u_{47} \over
u_{14} u_{15} u_{58}} \right) \nonumber\\
&& - {1\over8} \sum_{i=1}^8 \log^2 u_{i,i+3} - {1\over8}
\sum_{i=1}^8 [ 2 Li_2 \left( 1- u_{i,i+3} \right) + Li_2 \left( 1-
u_{i,i+4} \right) ] ~, \label{bds-8} \\
A_{\rm extra}^{n=8} & = & -{1\over 4}\left( \Delta_x \log
T_{2,2}^{[-3]} - \Delta_y \log T_{2,2}^{[-5]} \right) ~,
\label{extra-8} \eea
where $\Delta_{x,y}$ are given in (\ref{deltaxy}). We can see the
difference between BDS-like and BDS parts is indeed only a function
of cross ratios. The cross ratios in the large mass limit are
\bea && \{ u_{14}(i\phi) , ~  u_{25}(i\phi) , ~ u_{36}(i\phi) , ~
u_{47}(i\phi) , ~ u_{58}(i\phi) , ~
 u_{16}(i\phi) , ~  u_{27}(i\phi), ~  u_{38}(i\phi) \} \rightarrow \nonumber\\ && \qquad
\{ e^{-\sqrt{2}m_3 \cos(\phi-\pi/4)}, ~ e^{-\sqrt{2}m_3
\cos(\phi+\pi/4)} , ~ 1 ,~ 1 , ~ e^{-\sqrt{2}m_1 \cos(\phi-\pi/4)},
~ e^{-\sqrt{2}m_1 \cos(\phi+\pi/4)} , ~1, ~ 1 \} ~, \qquad \nonumber\\
&& \{ u_{15}(i\phi) , ~ u_{26}(i\phi) , ~ u_{37}(i\phi) , ~
u_{48}(i\phi) \} \rightarrow \{ e^{-\sqrt{2}m_2 \cos(\phi)} , ~ 1 ,
~1, ~ e^{-\sqrt{2}m_2 \sin(\phi)} \} ~. \label{8pointu} \eea
$\Delta_{x,y}$ are two special cross ratios that are related to the
monodromy \cite{Yang:2010az}, and can be given exactly as
\bea \Delta_x(\phi) &=& -\sqrt{2} (m_1 + \sqrt{2}m_2 +
m_3)\cos(\phi+\pi/4) ~, \\ \Delta_y(\phi) &=& -\sqrt{2} (m_1 +
\sqrt{2}m_2 + m_3)\sin(\phi+\pi/4) ~. \label{deltaxylimit} \eea
We also have extra part which involves $T$ functions. In the same
way as that for $Y$ function, we can obtain that
\bea \log T_{2,2}^{[-3]}(\phi) &\rightarrow& \sqrt{2} m_1 \cos(\phi+\pi/4) ~, \\
 \log T_{2,2}^{[-5]}(\phi) &\rightarrow& (2 m_1 + \sqrt{2}m_2 +
m_3) \cos(\phi) - m_3 \sin(\phi) ~. \eea

By substituting (\ref{8pointu})-(\ref{deltaxylimit}) into
(\ref{bds-8}) and (\ref{extra-8}), we obtain
\bea (A_{\rm BDS-like}^{n=8} - A_{\rm BDS}^{n=8})(\phi)
&\rightarrow& - {\sqrt{2}m_2\over 8}(m_1 + \sqrt{2}m_2 + m_3)(1 +
\sin(2\phi)- \cos(2\phi)) \nonumber\\ &&  - {1\over 4}(m_1 +
\sqrt{2}m_2 + m_3)^2 \cos(2\phi) - {m_1\over 2}(m_1 + \sqrt{2}m_2 +
m_3) \nonumber\\ && - {m_2^2 + m_3^2 + \sqrt{2}m_2 m_3 \over 2} +
{1\over 4}(m_1 + \sqrt{2}m_2 + m_3)^2 ~,
\\  A_{\rm extra}^{n=8}(\phi) &\rightarrow&
{\sqrt{2}m_2\over 8}(m_1 + \sqrt{2}m_2 + m_3)(1 + \sin(2\phi)-
\cos(2\phi)) \nonumber\\ && + {1\over 4}(m_1 + \sqrt{2}m_2 + m_3)^2
\cos(2\phi) + {m_1\over 2}(m_1 + \sqrt{2}m_2 + m_3) ~. \qquad \eea
We can see that $(A_{\rm BDS-like}^{n=8} - A_{\rm BDS}^{n=8})$ and
$A_{\rm extra}$ both depend on $\phi$ in a non-trivial way. However,
gratifyingly, the summation of them is independent of $\phi$, and we
have
\bea (A_{\rm BDS-like}^{n=8} - A_{\rm BDS}^{n=8})(\phi) + A_{\rm
extra}^{n=8}(\phi) &\rightarrow& - {m_2^2 + m_3^2 + \sqrt{2}m_2 m_3
\over 2} + {1\over 4}(m_1 + \sqrt{2}m_2 + m_3)^2 ~.  \eea
This non-trivial cancellation for the $\phi$-dependent terms is a
strong check for the result. The periods part is then given by
(\ref{constraint}) as
\bea A_{\rm periods}^{n=8} & = & {|m_2|^2 + |m_3|^2 \over 2} + {m_2
\bar m_3 + \bar m_2 m_3 \over 2 \sqrt{2}} - {1\over 4}|m_1 +
\sqrt{2}m_2 + m_3|^2 ~,\eea
where we have written it for general complex $m_s$ by using the real
condition. Comparing with the periods part conjectured in
\cite{Yang:2010az}, we can see we need to add a new term
\bea - {1\over 4}|m_1 + \sqrt{2}m_2 + m_3|^2 = -{1\over 4} |w_0|^2
~, \eea
where $w_0 \equiv m_1 + \sqrt{2}m_2 + m_3$ is the monodromy
\cite{Alday:2010ku,Yang:2010az}. Structurally, one may also write
the periods of eight-point as
\bea A_{\rm periods}^{n=8} & = & A_{\rm periods}^{n=7}\big|_{m_i
\rightarrow m_{i+1}} - {1\over 4}|w_0|^2 ~. \eea
%

We also calculate the twelve-points, and as well find the
non-trivial cancellation for $\phi$ dependent terms. The periods
part also shows interesting structure as
\bea A_{\rm periods}^{n=12} & = & A_{\rm periods}^{n=11}\big|_{m_i
\rightarrow m_{i+1}}  +  A_{\rm periods}^{n=10}\big|_{m_i
\rightarrow m_{i+1}} - {1\over 4}|w_0^{n=12}|^2 ~, \eea
where each term on the right hand side may be found in
\cite{Alday:2010vh}. This also shows that the conjectured result in
\cite{Yang:2010az} is not a full answer, which includes only the
first term.

\section{Conclusion}

In this paper, we study a special collinear limit of the amplitudes
at strong coupling which in the $Y$ system corresponds to a limit
that all mass parameters are taken to be very large. This limit
imposes a strong restriction on the amplitudes by which the periods
parts can be uniquely fixed by the already known other parts of the
amplitudes. This is particularly important for the case where the
number of gluons is a multiple of four, since the results of periods
part have not been calculated before. We do the calculation for
amplitudes up to twelve points. For the $n\neq 4K$ cases, we all
reproduce the known results, which provides a strong consistency
check for the method. We calculate the periods part for eight and
twelve points for the first time. The non-trivial cancellation of
the $\phi$-dependent terms shows that the calculations are
consistent. We also study some technical issues in detail which are
involved in the calculation, such as the relation between
traditional cross ratios and $Y$ function and their large mass
limit. The pole contributions due to large phase shift of the
spectral parameter are also discussed and the explicit integral
equations of $Y$ and $T$ functions are given. These relations can be
applied not only in the large mass limit considered in this paper,
but also in more general study of the amplitudes.

\section*{Acknowledgements}

The author has benefited from the discussion with L. Alday, A.
Brandhuber, E. Buchbinder, A. Sever, and G. Travaglini. He is
grateful to Brandhuber and Travaglini for the suggestion and
encouragement. He would like to thank the invitation of Imperial
college where this work was presented. The author would also like to
thank the anonymous referee for his or her careful reading of the
manuscript, and the valuable comments and suggestions for the
improvements. This work is supported by STFC.

\appendix

\section{Explicit relations between cross ratios from $Y$ functions \label{Y-cr}}

We write cross ratios that are commonly used in weak coupling
calculations in terms of $Y$ functions explicitly.

\subsection{Six-point}

For six-point case, we have
\bea u_{14} \equiv {x_{15}^2 x_{24}^2 \over x_{14}^2 x_{25}^2} ~,
\qquad u_{25} \equiv {x_{26}^2 x_{35}^2 \over x_{25}^2 x_{36}^2} ~,
\qquad u_{36} \equiv {x_{13}^2 x_{46}^2 \over x_{36}^2 x_{14}^2} ~.
\eea
Using $Y$ functions we obtain a one parameter family of cross ratios
\bea u_{36} = {Y_{2,1}^{[-1]}\over 1+Y_{2,1}^{[-1]}} ~, \qquad
u_{14} = {Y^+_{2,1}\over 1+Y^+_{2,1}} ~, \qquad u_{25} =
{Y^{[3]}_{2,1}\over 1+Y^{[3]}_{2,1}} ~. \eea
%

\subsection{Seven-point}

For seven-point case, we have six independent cross ratios. In the
practical calculation one can focus on seven consecutive-cusp cross
ratios,
\bea u_{i,i+3} \equiv {x_{i,i+2}^2 x_{i+3, i+6}^2 \over x_{i,i+3}^2
x_{i+2, i+6}^2} ~, \qquad i = 1,...,7~. \eea
Other cross ratios can be constructed from them directly. They are
related to $Y$ functions as
\bea && u_{37} = {Y^{[-2]}_{2,2}\over 1+Y^{[-2]}_{2,2}} ~, \qquad
u_{14} = {Y_{2,2}\over 1+Y_{2,2}} ~, \qquad u_{25} =
{Y^{[2]}_{2,2}\over 1+Y^{[2]}_{2,2}} ~,  \nonumber\\ &&  u_{36} =
{Y^{[-3]}_{2,1}\over 1+Y^{[-3]}_{2,1}} ~, \qquad u_{47} =
{Y^{-}_{2,1}\over 1+Y^{-}_{2,1}} ~, \qquad u_{15} =
{Y^{+}_{2,1}\over 1+Y^{+}_{2,1}} ~, \qquad u_{26} =
{Y^{[3]}_{2,1}\over 1+Y^{[3]}_{2,1}} ~. \qquad\qquad \eea
We mention that there are many other choice to define the cross
ratios, for example we also have
\bea && u_{37} = {Y^{[5]}_{2,1}\over 1+Y^{[5]}_{2,1}} ~, \qquad
u_{14} = {Y^{[7]}_{2,1}\over 1+Y^{[7]}_{2,1}} ~, \qquad u_{25} =
{Y^{[9]}_{2,1}\over 1+Y^{[9]}_{2,1}} ~. \eea
Different definitions are equivalent with each other, due to the
relation between $Y$ functions, in particular the periodic relation
of $Y$ functions
\bea Y_{2,1}^{[l]} = Y_{2,2}^{[l+7]} ~, \qquad Y_{2,s}^{[l]} =
Y_{2,s}^{[l+14]} ~. \eea
%

\subsection{Eight-point}

For eight-point case, we have nine independent cross ratios. We will
need the following twelve cross ratios in the calculation
\bea && u_{i,i+3} \equiv {x_{i,i+4}^2 x_{i+1, i+3}^2 \over
x_{i,i+3}^2 x_{i+1, i+4}^2} ~, \qquad i = 1,...,8~, \\ && u_{i,i+4}
\equiv {x_{i,i+5}^2 x_{i+1, i+4}^2 \over x_{i,i+4}^2 x_{i+1, i+5}^2}
~, \qquad i = 1,...,4~. \eea
They are related to $Y$ functions as
\bea && u_{48} = {Y^{[-2]}_{2,2}\over 1+Y^{[-2]}_{2,2}} ~, \quad
u_{15} = {Y_{2,2}\over 1+Y_{2,2}} ~, \quad u_{26} =
{Y^{[2]}_{2,2}\over 1+Y^{[2]}_{2,2}} ~, \quad u_{37} =
{Y^{[4]}_{2,2}\over 1+Y^{[4]}_{2,2}} ~, \nonumber \\
&& u_{47} = {Y^{[-3]}_{2,1}\over 1+Y^{[-3]}_{2,1}} ~, \quad u_{58} =
{Y^{-}_{2,1}\over 1+Y^{-}_{2,1}} ~, \quad  u_{16} =
{Y^{+}_{2,1}\over 1+Y^{+}_{2,1}} ~, \quad u_{27} =
{Y^{[3]}_{2,1}\over 1+Y^{[3]}_{2,1}} ~, \nonumber \\
&& u_{38} = {Y^{[-3]}_{2,3}\over 1+Y^{[-3]}_{2,3}} ~ , \quad u_{14}
= {Y^{-}_{2,3}\over 1+Y^{-}_{2,1}} ~, \quad u_{25} =
{Y^{+}_{2,3}\over 1+Y^{+}_{2,3}} ~, \quad u_{36} =
{Y^{[3]}_{2,3}\over 1+Y^{[3]}_{2,3}} ~. \qquad\qquad \eea
In this case, we have that
\bea Y_{2,1}^{[l]} = Y_{2,3}^{[l+8]} ~, \qquad Y_{2,2}^{[l]} =
Y_{2,2}^{[l+8]} ~, \qquad Y_{2,s}^{[l]} = Y_{2,s}^{[l+16]} ~. \eea
%

\subsection{Nine-point}

For nine-point case, we have twelve independent cross ratios. We
will need the following 18 cross ratios,
\bea && u_{i,i+3} \equiv {x_{i,i+4}^2 x_{i+1, i+3}^2 \over
x_{i,i+3}^2 x_{i+1, i+4}^2} ~, \qquad i = 1,...,9~, \\ && u_{i,i+4}
\equiv {x_{i,i+5}^2 x_{i+1, i+4}^2 \over x_{i,i+4}^2 x_{i+1, i+5}^2}
~, \qquad i = 1,...,9~. \eea
They are related to $Y$ functions as
\bea && u_{48} = {Y^{[-4]}_{2,2}\over 1+Y^{[-4]}_{2,2}} ~, ~~ u_{59}
= {Y^{[-2]}_{2,2}\over 1+Y^{[-2]}_{2,2}} ~, ~~ u_{16} =
{Y_{2,2}\over 1+Y_{2,2}} ~, ~~ u_{27} = {Y^{[2]}_{2,2}\over
1+Y^{[2]}_{2,2}} ~, ~~ u_{38} = {Y^{[4]}_{2,2}\over 1+Y^{[4]}_{2,2}}
~, \qquad \nonumber\\
&& u_{39} = {Y^{[-4]}_{2,4}\over 1+Y^{[-4]}_{2,4}} ~, ~~ u_{14} =
{Y^{[-2]}_{2,4}\over 1+Y^{[-2]}_{2,4}} ~, ~~ u_{25} = {Y_{2,4}\over
1+Y_{2,4}} ~, ~~ u_{36} = {Y^{[2]}_{2,4}\over 1+Y^{[2]}_{2,4}} ~, ~~
u_{47} = {Y^{[4]}_{2,4}\over 1+Y^{[4]}_{2,4}} ~, \nonumber\\
&& u_{58} = {Y^{[-3]}_{2,1}\over 1+Y^{[-3]}_{2,1}} ~, ~~ u_{69} =
{Y^{-}_{2,1}\over 1+Y^{-}_{2,1}} ~, ~~~ u_{17} = {Y^{+}_{2,1}\over
1+Y^{+}_{2,1}} ~, ~~ u_{28} = {Y^{[3]}_{2,1}\over
1+Y^{[3]}_{2,1}} ~, \nonumber \\
&& u_{49} = {Y^{[-3]}_{2,3}\over 1+Y^{[-3]}_{2,3}} ~, ~~ u_{15} =
{Y^{-}_{2,3}\over 1+Y^{-}_{2,3}} ~, ~~~ u_{26} = {Y^{+}_{2,3}\over
1+Y^{+}_{2,3}} ~, ~~ u_{37} = {Y^{[3]}_{2,3}\over 1+Y^{[3]}_{2,3}}
~. \eea
In this case, we have that
\bea Y_{2,1}^{[l]} = Y_{2,4}^{[l+9]} ~, \qquad Y_{2,2}^{[l]} =
Y_{2,3}^{[l+9]} ~, \qquad Y_{2,s}^{[l]} = Y_{2,s}^{[l+18]} ~. \eea

\section{$Y$ and $T$ functions in different phase regions\label{Y-phase}}

To obtain the traditional cross ratios, we need to consider a set of
$Y$ functions in deferent phase regions, $Y_{a,s}^{[k]}(i\phi)$. We
need to modify the integral equations and add pole contributions
from the kernels while crossing the lines of $\pm i\pi/4, \pm
i\pi/2$ and so on. In this appendix, we give some explicit formulas
which are used in the calculation of this paper. We take $m_s$ to be
real positive, and ${\rm Im}(\theta) \in (0, \pi/4)$. We use the
convention that
\bea f^{[l]}(\theta) = f(\theta +i l\pi/4 ) ~. \eea

We first define functions for all range of $\theta$
\bea F_{2,s}(\theta) & \equiv & - \sqrt{2} m_s \cosh(\theta) - K_2
\star \alpha_s - K_1 \star \beta_s ~, \\ F_{1,s}(\theta) & \equiv &
- m_s \cosh(\theta) - C_s - {1\over2} K_2 \star \beta_s - K_1 \star
\alpha_s - {1\over2} K_3 \star \gamma_s ~, \\
F_{3,s}(\theta) & \equiv & - m_s \cosh(\theta) + C_s - {1\over2} K_2
\star \beta_s - K_1 \star \alpha_s + {1\over2} K_3 \star \gamma_s ~.
\eea
where $\alpha, \beta, \gamma$ and kernels are given in
(\ref{abgamma}), (\ref{kernel}). We emphasize that the $Y$ functions
with a given integral form are only defined in a given phase region
of $\theta$. The $Y$ functions in general phase regions can be given
by including various pole contributions as
\bea \log Y_{2,s}(\theta) &=& F_{2,s}(\theta) ~,
\nonumber \\  \log Y_{2,s}^{[1]}(\theta) &=& F_{2,s}^{[1]}(\theta) -
\alpha_s(\theta) ~,
\nonumber\\ \log Y_{2,s}^{[2]}(\theta) &=& F_{2,s}^{[2]}(\theta) -
\alpha_s^{[1]}(\theta) - \beta_s(\theta) ~,
\nonumber\\ \log Y_{2,s}^{[3]}(\theta) &=& F_{2,s}^{[3]}(\theta) -
\alpha_s^{[2]}(\theta) - \beta_s^{[1]}(\theta) - \alpha_s(\theta) ~,
\nonumber\\ \log Y_{2,s}^{[4]}(\theta) &=& F_{2,s}^{[4]}(\theta) -
\alpha_s^{[3]}(\theta) - \beta_s^{[2]}(\theta) -
\alpha_s^{[1]}(\theta) ~,
\nonumber \\  \log Y_{2,s}^{[-1]}(\theta) &=& F_{2,s}^{[-1]}(\theta)
~,
\nonumber \\ \log Y_{2,s}^{[-2]}(\theta) &=& F_{2,s}^{[-2]}(\theta)
- \alpha_s^{[-1]}(\theta) ~,
\nonumber\\ \log Y_{2,s}^{[-3]}(\theta) &=& F_{2,s}^{[-3]}(\theta) -
\alpha_s^{[-2]}(\theta) - \beta_s^{[-1]}(\theta) ~,
\nonumber\\ \log Y_{2,s}^{[-4]}(\theta) &=& F_{2,s}^{[-4]}(\theta) -
\alpha_s^{[-3]}(\theta) - \beta_s^{[-2]}(\theta) -
\alpha_s^{[-1]}(\theta) ~, \hskip 5.6cm\eea
\bea \log Y_{1,s}(\theta) &=& F_{1,s}(\theta) ~,
\nonumber \\  \log Y_{1,s}^{[1]}(\theta) &=& F_{1,s}^{[1]}(\theta) -
{1\over2}\beta_s(\theta) + {1\over2}\gamma_s(\theta) ~,
\nonumber\\ \log Y_{1,s}^{[2]}(\theta) &=& F_{1,s}^{[2]}(\theta) -
{1\over2}\beta_s^{[1]}(\theta) + {1\over2}\gamma_s^{[1]}(\theta) -
\alpha_s(\theta) ~,
\nonumber\\ \log Y_{1,s}^{[3]}(\theta) &=& F_{1,s}^{[3]}(\theta) -
{1\over2}\beta_s^{[2]}(\theta) + {1\over2}\gamma_s^{[2]}(\theta) -
\alpha_s^{[1]}(\theta)  - {1\over2}\beta_s(\theta) +
{1\over2}\gamma_s(\theta) ~,
\nonumber\\ \log Y_{1,s}^{[4]}(\theta) &=& F_{1,s}^{[4]}(\theta) -
{1\over2}\beta_s^{[3]}(\theta) + {1\over2}\gamma_s^{[3]}(\theta) -
\alpha_s^{[2]}(\theta)  - {1\over2}\beta_s^{[1]}(\theta) +
{1\over2}\gamma_s^{[1]}(\theta) ~,
\nonumber \\  \log Y_{1,s}^{[-1]}(\theta) &=& F_{1,s}^{[-1]}(\theta)
~,
\nonumber \\  \log Y_{1,s}^{[-2]}(\theta) &=& F_{1,s}^{[-2]}(\theta)
- {1\over2}\beta_s^{[-1]}(\theta) - {1\over2}\gamma_s^{[-1]}(\theta)
~,
\nonumber\\ \log Y_{1,s}^{[-3]}(\theta) &=& F_{1,s}^{[-3]}(\theta) -
{1\over2}\beta_s^{[-2]}(\theta) - {1\over2}\gamma_s^{[-2]}(\theta) -
\alpha_s^{[-1]}(\theta) ~,
\nonumber\\ \log Y_{1,s}^{[-4]}(\theta) &=& F_{1,s}^{[-4]}(\theta) -
{1\over2}\beta_s^{[-3]}(\theta) - {1\over2}\gamma_s^{[-3]}(\theta) -
\alpha_s^{[-2]}(\theta)  - {1\over2}\beta_s^{[-1]}(\theta) -
{1\over2}\gamma_s^{[-1]}(\theta) ~, \qquad \eea
and similar for $Y_{3,s}^{[k]}$ with only the sign for the
$\gamma_s^{[r]}$ terms changed compared to $Y_{1,s}^{[k]}$.

\vskip 0.3cm

For the calculation of extra parts, we also list the formulas of
$T_{2,1}$ and $T_{2,2}$ that are given in (\ref{T21}) and
(\ref{T22}). Similarly, we first define functions for all range of
$\theta$
\bea G_{2,1}(\theta) & \equiv & K_2 \star \hat \beta_1 +K_1 \star
\hat \alpha_1 ~, \\  G_{2,2}(\theta) & \equiv & - \sqrt{2} m_1
\cosh(\theta) + K_2 \star \hat \beta_2 + K_1 \star \hat \alpha_2 ~,
\eea
where
\bea \hat \beta_s \equiv \log (1+ Y_{2,s}) ~, \qquad \hat \alpha_s
\equiv \log { (1+Y_{1,s})(1+Y_{3,s})} ~. \eea
Then we have $T$ functions in general phase regions as
\bea \log T_{2,s}(\theta) &=& G_{2,s}(\theta) ~,
\nonumber \\  \log T_{2,s}^{[1]}(\theta) &=& G_{2,s}^{[1]}(\theta) -
\hat \beta_s(\theta) ~,
\nonumber\\ \log T_{2,s}^{[2]}(\theta) &=& G_{2,s}^{[2]}(\theta) -
\hat \beta_s^{[1]}(\theta) - \hat \alpha_s(\theta) ~,
\nonumber\\ \log T_{2,s}^{[3]}(\theta) &=& G_{2,s}^{[3]}(\theta) -
\hat \beta_s^{[2]}(\theta) - \hat \alpha_s^{[1]}(\theta) - \hat
\beta_s(\theta) ~,
\nonumber\\ \log T_{2,s}^{[4]}(\theta) &=& G_{2,s}^{[4]}(\theta) -
\hat \beta_s^{[3]}(\theta) - \hat \alpha_s^{[2]}(\theta) - \hat
\beta_s^{[1]}(\theta) ~,
\nonumber \\  \log T_{2,s}^{[-1]}(\theta) &=& G_{2,s}^{[-1]}(\theta)
~,
\nonumber \\ \log T_{2,s}^{[-2]}(\theta) &=& G_{2,s}^{[-2]}(\theta)
- \hat \beta_s^{[-1]}(\theta) ~,
\nonumber\\ \log T_{2,s}^{[-3]}(\theta) &=& G_{2,s}^{[-3]}(\theta) -
\hat \beta_s^{[-2]}(\theta) - \hat \alpha_s^{[-1]}(\theta) ~,
\nonumber\\ \log T_{2,s}^{[-4]}(\theta) &=& G_{2,s}^{[-4]}(\theta) -
\hat \beta_s^{[-3]}(\theta) - \hat \alpha_s^{[-2]}(\theta) - \hat
\beta_s^{[-1]}(\theta) ~. \eea
%


\section{Results of amplitudes and remainder functions \label{results}}

We list the explicit results of amplitudes up to nine points in this
appendix. $A_{\rm BDS}$ is the one-loop finite part of amplitudes at
weak couping \cite{BDS}. The result of $A_{\rm BDS-like}$, $A_{\rm
periods}$ and $A_{\rm free}$ for $n\!\neq\! 4K$ cases were given in
\cite{Alday:2010vh}. $A_{\rm extra}$ and $A_{\rm BDS-like}$ for
eight-point were given in \cite{Yang:2010az}. The periods part of
eight-point are calculated in this paper. We also give the result of
the difference between BDS-like and BDS part in terms of cross
ratios beyond six-point for the first time. The total amplitudes and
remainder functions can be constructed as
\bea A  & = &  A_{\rm div} + A_{\rm BDS-like} + A_{\rm extra} +
A_{\rm periods} + A_{\rm free} ~, \\ R  & = &  (A_{\rm BDS-like}
-A_{\rm BDS}) + A_{\rm extra} + A_{\rm periods} + A_{\rm free} ~.
\eea
The IR divergent parts are universal
\bea A_{\rm div} = {1\over 2} \sum_{i=1}^n \Big( L+{\ell_i \over2}
\Big)^2 ~, \eea
where $L$ is the cutoff for the minimal surface. We use the notation
that
\bea \ell_i \equiv \log x_{i,i+2}^2 ~, \qquad \ell_{ij} \equiv \log
x_{ij}^2 ~. \eea
%

\subsection{Six-point}
\bea && A_{\rm periods} = {|m_1|^2 \over 4} ~, \\ && A_{\rm
BDS-like} = -{1\over 8} \sum_{i=1}^6 \Big( \ell_i^2 + \sum_{k=0}^{2}
\ell_i \ell_{i+2k+1} (-1)^{k+1} \Big) ~, \\ &&  A_{\rm free} =
{|m_1| \over 2\pi} \int_{-\infty}^{+\infty} d\theta ~ \cosh\theta
\log \left[(1+ Y_{1,1})(1+ Y_{3,1})(1+ Y_{2,1})^{\sqrt{2}}\right]~.
\eea
\bea A_{\rm BDS-like} - A_{\rm BDS} = -\sum_{i=1}^3 \left(
{1\over8}\log^2 u_i + {1\over4} Li_2(1-u_i)\right) ~.  \eea
%

\subsection{Seven-point}

\bea && A_{\rm BDS-like} = -{1\over 4} \sum_{i=1}^7 \Big( \ell_i^2 +
\sum_{k=0}^{2} \ell_i \ell_{i+2k+1} (-1)^{k+1} \Big) ~, \\ && A_{\rm
periods} = {|m_1|^2 + |m_2|^2 \over 2} + {m_1 \bar m_2 + \bar m_1
m_2 \over 2 \sqrt{2}} ~, \\ && A_{\rm free} = \sum_{s=1}^2 { |m_s|
\over 2\pi} \int_{-\infty}^{+\infty} {d\theta } \cosh\theta \log
\left[(1+ Y_{1,s})(1+ Y_{3,s})(1+ Y_{2,s})^{\sqrt{2}}\right] ~. \eea
\bea A_{\rm BDS-like} - A_{\rm BDS} = -{1\over 4}\sum_{i=1}^7 \left(
\log^2 u_{i,i+3} + Li_2(1-u_{i,i+3}) -{1\over2} \log u_{i,i+3} \log
{u_{i+2,i+5} u_{i+1,i+5} \over u_{i+3,i+6} u_{i,i+4}} \right) ~.
\nonumber\\  \eea
%

\subsection{Eight-point}

\bea && A_{\rm periods} = {|m_2|^2 + |m_3|^2 \over 2} + {m_2 \bar
m_3 + \bar m_2 m_3 \over 2 \sqrt{2}}  - {1\over 4}|m_1 + \sqrt{2}m_2
+ m_3|^2~,
\\ && A_{\rm free} = \sum_{s=1}^3 { |m_s| \over 2\pi}
\int_{-\infty}^{+\infty} {d\theta }  \cosh\theta \log \left[(1+
Y_{1,s})(1+ Y_{3,s})(1+ Y_{2,s})^{\sqrt{2}}\right] ~, \\ && A_{\rm
BDS-like} = -{1\over 8} \sum_{i=1}^{8} \ell_i^2  + {1\over 4}
\sum_{i=1}^{8} \ell_i \ell_{i+1} - {1\over 4} (\ell_2 + \ell_6)
(\ell_3 + \ell_7) + {1\over 4}\left( \Delta_x \ell_{48}  - \Delta_y
\ell_{37} \right)  ~, \\ && A_{\rm extra} = -{1\over 4}\left(
\Delta_x \log T_{2,2}^{[-3]} - \Delta_y \log T_{2,2}^{[-5]} \right)
~, \eea
where
\bea \Delta_x = -\ell_1+\ell_3-\ell_5+\ell_7 ~, \qquad \Delta_y =
-\ell_2+\ell_4-\ell_6+\ell_8 ~. \label{deltaxy} \eea
\bea A_{\rm BDS-like} - A_{\rm BDS} &=& - {1\over8} \log^2 \left(
{u_{14} u_{25} u_{58} u_{16} \over u_{36} u_{47} u_{27}} \right) -
{1\over4} \log \left( u_{15} u_{16} u_{25} u_{26} \right) \log
\left( u_{26} u_{27} u_{36} u_{37} \right) \nonumber\\ && -
{1\over2} \log \left( {u_{15} \over u_{37}} \right) \log \left(
{u_{14} u_{25} u_{58} u_{16} u_{15} \over u_{36} u_{47} u_{27}
u_{37}} \right)
- {1\over8} (\Delta_x - \Delta_y)^2 \nonumber\\
&& + {1\over4} \Delta_x \log \left( {u_{14} u_{15}^2 u_{16} u_{25}
u_{58} \over u_{27} u_{36} u_{37}^2 u_{47}} \right) + {1\over4}
\Delta_y \log \left( {u_{26} u_{27} u_{36} u_{37}^2 u_{47} \over
u_{14} u_{15} u_{58}} \right) \nonumber\\
&& - {1\over8} \sum_{i=1}^8 \log^2 u_{i,i+3} - {1\over8}
\sum_{i=1}^8 [ 2 Li_2 \left( 1- u_{i,i+3} \right) + Li_2 \left( 1-
u_{i,i+4} \right) ] ~. \eea
%

\subsection{Nine-point}

\bea &&  A_{\rm BDS-like} = -{1\over 4} \sum_{i=1}^9 \Big( \ell_i^2
+ \sum_{k=0}^4 \ell_i \ell_{i+1+2k} (-1)^{k+1} \Big) ~,
\\ && A_{\rm free} = \sum_{s=1}^4 { |m_s| \over 2\pi}
\int_{-\infty}^{+\infty} {d\theta }  \cosh\theta \log \left[(1+
Y_{1,s})(1+ Y_{3,s})(1+ Y_{2,s})^{\sqrt{2}}\right] ~, \\ && A_{\rm
periods} = -{1\over2} (|m_2|^2 + |m_3|^2 + m_1 \bar m_3 + m_3 \bar
m_1 + m_2 \bar m_4 + m_4 \bar m_2 ) \\ && \qquad\qquad - {\sqrt{2}
\over 4} (m_1 \bar m_2 + m_2 \bar m_1 + m_1 \bar m_4 + m_4 \bar m_1
+ m_3 \bar m_4 + m_4 \bar m_3 + 2 m_2 \bar m_3 + 2 m_3 \bar m_2) ~.
\nonumber \eea
\bea A_{\rm BDS-like} - A_{\rm BDS} &=& {1\over 4}\sum_{i=1}^9
\left[ \log u_{i,i+3} \log \left( {u_{i,i+4} u_{i+4,i+7} \over
u_{i+1,i+5} u_{i+2,i+5} u_{i+3,i+6} u_{i+2,i+6}^2} \right) \right.
\\ && \hskip -2cm \left. + \log u_{i,i+4} \log \left(
{u_{i,i+4} u_{i+1,i+4} u_{i+4,i+8}^2 u_{i+5,i+8}^2 \over u_{i+2,i+5}
u_{i+2,i+6}^3 u_{i+3,i+6}^2 u_{i+3,i+7}^2} \right) - Li_2 \left( 1-
u_{i,i+3} \right) - Li_2 \left( 1- u_{i,i+4} \right) \right] ~.
\nonumber \eea
%


\end{document}